# Zur Modellierung und Klassifizierung von Kompetenzen in der grundlegenden Programmierausbildung anhand der Anderson Krathwohl Taxonomie[1]

Natalie Kiesler[2]

**Abstract:** In diesem Forschungsbeitrag werden Kompetenzanforderungen an Informatik-Studierende speziell während der grundlegenden Programmierausbildung betrachtet. Basierend auf einer qualitativen Inhaltsanalyse aktueller Lehr- und Lernziele deutscher Hochschulen und der anhand von Interviews erhobenen Perspektive von Hochschullehrenden werden angestrebte Programmierkompetenzen der Basisausbildung identifiziert. Da das Kompetenzmodell der Gesellschaft für Informatik diverse Defizite aufweist, werden die entwickelten inhaltlichen Kategorien in die Anderson Krathwohl Taxonomie (AKT) kognitiver Bildungsziele eingeordnet. Daraufhin erfolgt eine Überarbeitung der Dimensionen und Subtypen der AKT hin zu einem für die Informatik spezifischen Modell mit dem Ziel, Programmierkompetenzen gemäß ihrer kognitiven Komplexität klassifizieren zu können. Die präsentierte Handreichung kann auf den Ebenen der Lehrveranstaltungskonzeption, Entwicklung und Bewertung von Assessments genutzt werden. Weiterhin können die empirisch erhobenen und kategorisierten Kompetenzen entlang der AKT perspektivisch zur Konstruktion eines Instruments zur Messung von Programmierkompetenzen dienen.

## 1   Relevanz von Kompetenzmodellen in der Informatik

Als Konsequenz internationaler Vergleichsstudien wie dem *Programme for International Student Assessment* (PISA) [BSD01], entflammte in Deutschland die Debatte um Bildungsstandards und zentralisierte Abschlüsse [Wa09]. Durch bildungspolitische und -theoretische Diskurse rückten *Kompetenzen* und *Kompetenzmodelle* in den Vordergrund. Weinert definiert *Kompetenz* als individuelle Dispositionen erlernbarer Fähigkeiten zur Problemlösung in immer wieder neuen, unbekannten Situationen. Es werden kognitive und metakognitive Kompetenzen differenziert, die jeweils als kontextspezifisch zu begreifen sind. Darüber hinaus werden durch Handlungskompetenzen motivationale und volitionale Aspekte ergänzt, die zur Bewältigung komplexer Aufgaben beitragen [We01a, We01b]. Klieme unterstreicht die Bedeutung individueller Performanz bei der Problemlösung. Die Kompetenzforschung versucht daher individuelle Ausprägungen von Kompetenzen zu untersuchen, indem diese eindeutig bestimmt, operationalisiert, d. h. sichtbar und dadurch messbar werden [KH07].

---

[1] Eine Kurzfassung dieses Beitrags wurde im Rahmen der DELFI 2020 veröffentlicht [Ki20a].
[2] Goethe-Universität Frankfurt, Akademie für Bildungsforschung und Lehrerbildung, Juridicum, Senckenberganlage 31-33, 60325 Frankfurt am Main, kiesler@em.uni-frankfurt.de

Dazu ist die Verknüpfung mit konkreten, fachspezifischen Inhalten notwendig. Kompetenzmodelle sind daher nicht zu verwechseln mit Auflistungen von Fachthemen. Vielmehr werden systematisch aufeinander aufbauende Kompetenzen eines Fachgebiets inhaltsbezogen, etwa in Form einer Matrix mit verschiedenen Dimensionen, abgebildet [Kl04]. Der Entwicklung kontextspezifischer Kompetenzmodelle wurde in der Informatik bisher wenig Beachtung geschenkt. Zum einen ist die Informatik eine sich stetig weiterentwickelnde Fachdisziplin, zum anderen benötigen valide Instrumente zur Kompetenzmessung empirische Daten als Basis [Ko08, KTB16b, KTB16a]. Zwar wurde in den vergangenen Jahren besonders in Deutschland an Bildungsstandards für den Primarbereich [Ar19] sowie für Sekundarstufe 1 und 2 [Ar08, Ar16] gearbeitet, die Entwicklung empirisch abgeleiteter Kompetenzmodelle für die universitäre Informatik-Ausbildung findet jedoch nur in vereinzelten Forschungsarbeiten Berücksichtigung [Li13, KTB16a, Sc12]. Durch die Gesellschaft für Informatik (GI) wurde zwar ein Kompetenzmodell als Empfehlung für die Entwicklung von Kompetenzen im Informatik-Studium entwickelt [Ge16]. Einige Wissensdimensionen sowie kognitive Prozessdimensionen gemäß der Anderson Krathwohl Taxonomie (AKT) [AK01] werden dabei allerdings nicht berücksichtigt bzw. aufgrund von Missverständnissen bezüglich Kontextspezifika der Informatik fehlinterpretiert.

In Anbetracht zunehmend heterogener Zielgruppen an den Hochschulen besteht auf verschiedenen Ebenen Handlungsbedarf, u. a. auf didaktischer Ebene, bei der Gestaltung von Curricula und bezüglich Assessments. Insbesondere in der grundlegenden Programmierausbildung als Kern jedes Informatik-Studiums [JS13] fallen seit jeher zahllose Herausforderungen für Novizen auf [SS86, Wi96, RRR03, BC07, GM07, Xi16]. Eine systematische Evaluation verschiedener Lehrkonzepte konnte dahingehend keine signifikanten Auswirkungen auf Erfolgsquoten in Lehrveranstaltungen für ProgrammieranfängerInnen feststellen [VAW14]. Die vorliegende Forschungsarbeit untersucht daher die Anwendbarkeit der AKT [AK01] als anerkannten Rahmen zur Kategorisierung kognitiver Kompetenzen in der grundlegenden Programmierausbildung. Durch die qualitative Inhaltsanalyse [Ma15] von Informatik-Curricula 35 deutscher Hochschulen sowie der qualitativen Auswertung Leitfaden-gestützter ExpertInneninterviews erfolgt eine Klassifizierung aktuell erforderlicher Kompetenzen im Bereich der grundlegenden Programmierausbildung. Die darauffolgende Einordnung in die Dimensionen der AKT zeigt deren Anwendbarkeit für die kontextspezifischen Anforderungen speziell in der Programmierausbildung. Diese Klassifizierung birgt das Potenzial, Programmierkompetenzen durch didaktische Maßnahmen schrittweise und operationalisiert aufbauen und folglich testen zu können. Ein solches Vorgehen kann Lehrende und Lernende mittels zunehmender Transparenz und vereinfachter Messbarkeit von Lernergebnisse unterstützen, sodass die Qualität der Lehre verbessert werden kann.

## 2 Kompetenzmodelle für der Informatik

Im Folgenden wird die AKT als Basis dieser Arbeit vorgestellt. Weiterhin wird das darauf aufbauende, von der Gesellschaft für Informatik (GI) entwickelte Kompetenzmodell für Bachelor- und Masterprogramme im Studienfach Informatik kritisch diskutiert.

### 2.1 Anderson und Krathwohl Taxonomie

Die Bloom'sche Taxonomie kognitiver Lehr- und Lernziele [Bl56] wurde im Jahr 2001 von Anderson und Krathwohl mit Blick auf die zunehmende Entwicklung von Bildungsstandards und deren Verwendung in Curricula revidiert [AK01]. Es wurden Schwerpunkte, Terminologie und Strukturelemente der Taxonomie angepasst, um Bildungsziele klassifizieren zu können. Das Ergebnis ist eine zweidimensionale Matrix (siehe Tabelle 1a) bestehend aus vier Wissensdimensionen (*Faktenwissen*, *Konzeptionelles Wissen*, *Prozedurales Wissen*, *Metakognitives Wissen*) und sechs kognitiven Prozessdimensionen (*Erinnern*, *Verstehen*, *Anwenden*, *Analysieren*, *Bewerten*, *Erzeugen*). Anhand der Dimensionen wird ein Kontinuum kognitiver Komplexität abgebildet. Niedrigere kognitive Prozessdimensionen wie etwa *Erinnern* und *Verstehen* werden vorausgesetzt, um Lernziele auf der Ebene des *Anwendens* meistern zu können. Die Wissensdimensionen sind ebenfalls entlang eines Kontinuums angelegt und reichen von konkretem Faktenwissen bis hin zu abstraktem, metakognitivem Wissen [AK01, p. 5]. Konzeptionelles und prozedurales Wissen kann sich dabei überschneiden, etwa weil prozedurales Wissen konkreter sein kann als abstraktes, konzeptionelles Wissen. Ein operationalisiertes Lernziel wird durch Verb und Substantiv gekennzeichnet. Das Verb beschreibt den angestrebten kognitiven Prozess, das verwendete Substantiv zu erwerbendes bzw. zu konstruierendes Wissen.

Tab. 1: Übersicht der referenzierten Modelle zur Klassifizierung von Kompetenzen

(a) Anderson Krathwohl Taxonomie [AK01]

| Knowledge Dimension | Cognitive Process Dimension | | | | | |
|---|---|---|---|---|---|---|
| | *Remember* | *Understand* | *Apply* | *Analyze* | *Evaluate* | *Create* |
| *Factual knowledge* | | | | | | |
| *Conceptual knowledge* | | | | | | |
| *Procedural knowledge* | | | | | | |
| *Metacognitive knowledge* | | | | | | |

(b) Kompetenzmodell der GI [Ge16]

| Kognitive Prozessdimension | Stufe 1 Verstehen | Stufe 2 Anwenden | Stufe 3 Analysieren | Stufe 4 Erzeugen |
|---|---|---|---|---|
| Geringe Kontextualisierung und Komplexität | | | | |
| | | Stufe 2a Übertragen | Stufe 3a Bewerten | |
| Starke Kontextualisierung und hohe Komplexität | | | | |

### 2.2 Kompetenzmodell der Gesellschaft für Informatik

Die GI veröffentlichte 2016 im Rahmen ihrer „Empfehlungen für Bachelor- und Masterprogramme im Studienfach Informatik an Hochschulen" [Ge16] ein Kompetenzmodell basierend auf der AKT [AK01]. Ziel dieses Modells ist es, die in Informatik-Studiengängen angestrebten kognitiven Kompetenzen zu beschreiben und damit Hilfestellung zur Gestaltung, Weiterentwicklung und Beurteilung derselben zu ermöglichen [Ge16]. Die Skizze des GI-Modells in Tabelle 1b differenziert in der ersten Zeile die vier kognitiven Prozessdimensionen *Verstehen*, *Anwenden*, *Analysieren* und *Erzeugen*. Durch die beiden Zeilen soll zwischen *Typen des Wissenschaftlichen Arbeitens* mit den Subtypen T1 bis T6 unterschieden werden, die durch *Stufen der Kontextualisierung von Anwendungen* (K1 bis K5) und Wissensdimensionen (W1 bis W4) charakterisiert werden. Die notwendigen Erläuterungen zur Bedeutung der Wissensdimensionen für die Informatik und zur reduzierten Darstellung in Form von zwei Zeilen fehlen jedoch. Implizierte Annotationen zu den jeweiligen Typen T, K und W in den Beispielen der 17 Inhaltsbereiche bleiben aus. Definierte Kategorien (K5) bleiben ungenutzt und die Entwicklung metakognitiver Kompetenzen scheint ausschließlich der Promotion vorbehalten zu sein. Indem die Klassifizierungen in den Beispielen von den Autoren selbst nicht präzise und nachvollziehbar angewendet werden können, offenbart sich eine der größten Schwächen des Modells. Lediglich kognitive Prozessdimensionen werden in den Beispielen zumindest erkennbar (wenn auch nicht nachvollziehbar) eingeordnet.

Die zweite Zeile des Modells beinhaltet weitere Stufen kognitiver Prozessdimensionen: 2a *Übertragen* und 3a *Bewerten*. Die aus der AKT bekannte Dimension *Erinnern* wird wegen vermeintlicher Redundanz nicht berücksichtigt. Die Felder in der zweiten Zeile unter Stufe 1 und 4 bleiben leer, weil „sich die Prozessdimension ‚Verstehen' eher auf grundlegende Zusammenhänge beziehen soll und Stufe 4 ‚Erzeugen' im Bachelor-Studiengang kaum grundlegende wissenschaftliche Innovation in komplexeren Anwendungszusammenhängen erwarten lässt" [Ge16, S. 10]. Demnach würde die Dimension *Erzeugen* in Bachelor-Studiengängen nicht angestrebt und bleibt ungenutzt. Spätestens an dieser Stelle wird die Fehleinschätzung zum kognitiven Anspruch allein der Programmierung deutlich.

Gegenläufig zum GI-Modell liefert die AKT bereits ein komplexes Kontinuum kognitiver Kompetenzen und Wissensdimensionen. Eine weitere, explizite Differenzierung nach Kontextualisierung von Inhalten durch die GI ist nicht notwendig, da die Wissensdimensionen kontextspezifisches Wissen anhand der Subtypen adressieren. Darüber hinaus fasst das GI an anderer Stelle Dimensionen zusammen, die zu differenzierende kognitive Prozesse abbilden, wie zum Beispiel *Erinnern* und *Verstehen*. Auch in der Informatik und speziell der Programmierung müssen Studierende grundlegende Fakten und Informationen aus dem Langzeitgedächtnis abrufen können, so z. B. Schlüsselworte, Operatoren, Escape-Sequenzen, Wertebereiche, etc. Das Erinnern an derartige Details ist ohne jegliches Verständnis möglich, sodass die Position der GI weiter konterkariert wird. Besonders scharfe Kritik bedarf die Position zur Kategorie des *Erzeugens*. Während die GI damit einhergehende kognitive Prozesse Bachelor- und Master-Arbeiten vorbehält, vertritt die Autorin die Position, dass bereits das Schreiben kleiner Algorithmen und Programme derartige planerische, gestalterische und konstruktive Leistungen zur Produktion neuer Lösungen erforderlich macht. Diese kognitiven Prozesse sind entsprechend der AKT in die Dimension des *Erzeugens* einzuordnen. Zusätzlich werden metakognitive Kompetenzen benötigt, um Lernprozesse zu organisieren und eine Systematik bei der Problemlösung in Form von Programmen entwickeln zu können. Laut GI werden metakognitiven Kompetenzen jedoch erst ab der Promotion ausgebildet [Ge16, S. 11]. Demnach lernen Studierende im Bachelor nicht die strategische Planung von Lernprozessen und erreichen keine Selbsterkenntnis.

Da auch die weiteren Einordnungen der GI entlang der verbleibenden kognitiven Prozessdimensionen nicht den Typen, Subtypen und Kategorien kognitiver Prozesse entsprechen, bedarf es einer umfassenden neuen Handreichung für die Klassifizierung von Kompetenzen entlang der AKT-Matrix in der Informatik. Besonders die Programmierung kann von einem komplexeren Kompetenzmodell unter Berufung auf die AKT profitieren, um die Gestaltung von Lehrangeboten für AnfängerInnen zu optimieren, da bereits alle kognitiven Prozesse und Wissensdimensionen darin abgebildet werden. Dadurch kann zum einen bei Lehrenden das Bewusstsein für den kognitiven Anspruch der Programmierung geschärft, zum anderen können Aufgaben entlang der verschiedenen Dimensionen entwickelt und damit lernförderlicher gestaltet werden. Nicht zuletzt können (Self-)Assessments Programmierkompetenzen differenzierter abbilden und Defizite im Wissensstand klarer identifizieren.

## 3  Forschungsdesign

Die Anderson Krathwohl Taxonomie zur Klassifikation von Bildungszielen bildet kognitive Prozesse und Wissensdimensionen in ausreichender Komplexität und unter Berücksichtigung von Fachspezifika ab. Eine Reduktion in Kombination mit neuen Dimensionen zur Kontextualisierung, wie im GI-Modell vorliegend, erscheint paradox. Aufgrund der Fehlinterpretationen der GI wird die AKT vor dem Hintergrund der kognitiven Anforderungen in der Programmierung als essenzieller Kern jedes Informatik-Studiums neu bewertet. Dazu werden folgende Forschungsfragen untersucht: *(1) Inwieweit kann die Anderson Krathwohl Taxonomie in der Grundlagenausbildung der Programmierung Anwendung finden, um Programmierkompetenzen zu klassifizieren? (2) Wie fachspezifisch ist die AKT zu begreifen?*

Auf Grundlage der ACM-Curricula Empfehlungen [JS13] zu typischen Inhalten in Informatik-Studiengängen wurde ein Inhaltsbereich für die grundlegende Programmierausbildung in den ersten drei bis vier Semestern (abhängig vom Studienstart zum Sommer- oder Wintersemester) eingegrenzt. Anhand des Inhaltsbereichs wurden von je einer zufällig ausgewählten Universität und Hochschule pro Bundesland reine Informatik-Studiengänge analysiert, die mit Bachelor of Science abschließen. Ausgenommen wurden alternative Formate wie duale und Teilzeit-Studiengänge, Informatik als Nebenfach, die Bindestrich-Informatik und Angebote von Privathochschulen. Außerdem werden dem Sample die Bildungseinrichtungen der interviewten ExpertInnen hinzugefügt, die nicht schon zufällig im Sample integriert waren, sodass insgesamt 35 Institutionen betrachtet wurden. Aus 129, dem Inhaltsbereich entsprechenden Modulen (z. B. „Programmierung 1", „Informatik 1", „Algorithmen und Datenstrukturen") wurden Lernziele extrahiert und anhand der qualitativen Inhaltsanalyse nach Mayring [Ma15] in MAXQDA kodiert. Die deduktiv-induktiv gebildeten, abstrahierten Kategorien der gleichnamigen inhaltsbezogenen Kompetenzen wurden zweifach kodiert. Zum einen entsprechend der Wissensdimension und zum anderen gemäß deren kognitiver Prozessdimension in der AKT. Nicht-kognitive Kompetenzen sind gleichnamig kodiert.

Anhand von sieben Leitfaden-gestützte Interviews mit Hochschullehrenden als ExpertInnen [Fr97] der Programmierausbildung konnten weitere Programmierkompetenzen sowie Herausforderungen für Lernende identifiziert werden. ProfessorInnen als ExpertInnen mit langjähriger Lehrerfahrung kennen typische Herausforderungen von ProgrammiernovizInnen und können strukturelle Schwierigkeiten offen reflektieren. So werden Einblicke in spezialisiertes Wissen und Erfahrungen der jeweiligen Einrichtungen möglich, die trotz subjektiver Wahrnehmungen auf andere Bildungsinstitutionen übertragen werden können [Ku08]. Nach der Recherche potentieller ExpertInnen und schriftlicher Kontaktaufnahme konnten durch das Schneeball-Prinzip weitere befragte Personen für etwa einstündige Interviews gewonnen werden. Dabei wurden unter anderem folgende, den interviewten Personen unbekannte Fragen gestellt, die jeweils durch weitere Nachfragen ergänzt wurden.

1. Was bedeutet es für Sie, programmieren zu können?
2. Was fällt Ihnen immer wieder in Ihren Lehrveranstaltungen auf in Bezug auf das Erlernen von Programmierfähigkeiten?
3. Welche Faktoren bedingen den Erwerb von Programmierfertigkeiten?

Die Interviews wurden durch einen kurzen Fragebogen und ausführliche Notizen, respektive eine Audioaufzeichnung begleitet, die später transkribiert wurde [KO09]. Alle Transkripte wurden anhand einer zusammenfassenden qualitativen Inhaltsanalyse nach Mayring [Ma15] auf beschriebene Programmierkompetenzen hin untersucht. Dementsprechend wurde das Material durch Paraphrasieren, Generalisieren, zwei Stufen der Reduktion und weitere Makrooperatoren zusammengefasst [Ba81], bevor Kategorien kodiert werden konnten. In Abgleich mit den im Zuge der Curricula-Analyse entwickelten kognitiven Kategorien wurden ggf. weitere, induktive Kategorien gebildet, und ebenfalls in die Dimensionen der AKT eingegliedert. Darüber hinaus wurden nicht-kognitive Kompetenzen identifiziert.

## 4  Ergebnisse

Als Ergebnis der beiden qualitativen Analysen wurden die angestrebten Kompetenzen der grundlegenden Programmierausbildung während der ersten drei, respektive vier Semester an deutschen Hochschulen entsprechend der Wissensdimensionen und kognitiven Prozessdimensionen der AKT tabellarisch zusammengefasst. Nicht-kognitive Kompetenzen wurden in Listenform ergänzt.[3] Die Klassifikationen der beiden Samples wurden zusammengeführt und entsprechend des festgelegten Inhaltsbereichs um einige Ziele bereinigt, sodass der Umgang mit LaTeX, oder Konzepte aus der theoretischen Informatik beispielsweise entfernt wurden.[4] Die entstandene Übersicht der normativen und empirisch erhobenen Programmierkompetenzen zeigt die grundsätzliche Anwendbarkeit der AKT für die Basisausbildung der Programmierung sehr deutlich. Wie vermutet, stellt sich eine ausgewogene Verteilung entlang der Diagonale der AKT ein. Weiterhin bestätigt das Datenmaterial die Anordnung kognitiver Prozesse aus der revidierten Taxonomie [AK01] im Gegensatz zur Bloom'schen Taxonomie [Bl56]. Die analysierten Bildungsziele der kognitiven Prozessdimensionen beschreiben die Ebenen *Erzeugen* und *Bewerten* in einer Art und Weise, die den stufenweisen Anstieg kognitiver Komplexität bis hin zur höchsten Ebene des *Erzeugens* widerspiegelt.

---

[3] Die Datengrundlage der vorliegenden Arbeit wird durch das jeweilige Codebuch und die Übersicht codierter Segmente zur Verfügung gestellt: https://github.com/nkiesler-cs/delfi2020.git

[4] Die Klassifizierung aller kognitiver Kompetenzen sowie die vollständige Auflistung nicht-kognitiver Kompetenzen wurde bereits hier veröffentlicht [Ki20b].

Um die Einordnung kognitiver Kompetenzen in die AKT zu illustrieren, werden einige Beispiele für den Inhaltsbereich der grundlegenden Programmierausbildung in Tabelle 2 angeführt. Dazu werden die exemplarischen Kompetenzen des Inhaltsbereichs Programmiersprachen und -methodik der GI [Ge16, S.31-32] präzisiert und anhand von Beispielen aus den empirisch gewonnenen Daten erweitert, um die kognitiven Prozessdimensionen typischer Programmierkompetenzen sichtbar zu machen. Die Achsen der AKT-Matrix wurden dabei lediglich aus Gründen der Formatierung transponiert.

Tab. 2: Exemplarische Anwendung der AKT für die grundlegende Programmierausbildung

| Kognitive Prozessdimension | Wissensdimensionen | | | |
|---|---|---|---|---|
| | *Faktenwissen* | *Konzeptionelles Wissen* | *Prozedurales Wissen* | *Metakognitives Wissen* |
| *Stufe 1 Erinnern* | Erinnern an Elemente einer Programmiersprache (Sprachkonstrukte wie z. B. Schlüsselworte, Literale, Operatoren) | Erinnern an Konzepte von Programmierparadigmen und -sprachen (z. B. Typsystem, Funktionen, Parameterübergabe, Speicherverwaltung) | Erinnern an konkrete Syntax und Semantik von Programmiersprachen, und Konventionen zum Programmierstil<br><br>Werkzeuge zur Softwareentwicklung, deren Funktionen und Anwendungszwecke kennen (Laufzeitumgebung, IDE) | |
| *Stufe 2 Verstehen* | | Konzepte von Programmierparadigmen und -sprachen illustrieren und vergleichen können | Unterschiede und Funktionsweise von Compiler und Interpreter erklären können<br><br>Merkmale von grundlegenden Datentypen, Datenstrukturen, Algorithmen und Entwurfsmustern erklären | |
| *Stufe 3 Anwenden* | | | Qualitätskriterien und Programmierkonventionen umsetzen<br><br>Zahlen in unterschiedliche Darstellungsformen übertragen und damit rechnen | |
| *Stufe 4 Analysieren* | | | Gegebene Problemstellungen in Teile zerlegen und die wichtigsten Informationen auswählen<br><br>Fremden Code zerlegen und dessen innere Struktur erkennen, sodass dessen Ausgabe(n) bestimmt werden können | |
| *Stufe 5 Bewerten* | | | Angemessenheit von Algorithmen (inkl. Datenstrukturen, Datentypen) und programmiersprachlichen Lösungen für eine Problemlösung beurteilen<br><br>Testen von Algorithmen und Programmen bzgl. ihrer Eigenschaften und auf Fehler | Überprüfen und Beurteilen des eigenen Lernhandelns (Selbstreflexion) |
| *Stufe 6 Erzeugen* | | | Problemadäquate Datenstrukturen und Algorithmen auswählen und unter Einsatz bekannter Techniken und Muster entwerfen<br><br>Programmiersprachliche, lauffähige Lösungen für kleinere bis mittlere Probleme schreiben (bis zu ein paar hundert Zeilen Code) | Systematisches Vorgehen beim Problemlösen entwickeln<br><br>Einarbeiten in neue Programmiersprache eines bekannten Paradigmas durch Transfer von vorhandenem Wissen und Erfahrung |

Insbesondere die ExpertInnen-Interviews offenbarten das Entwickeln und Schreiben von (wenigen Zeilen) Programmcode als Herausforderung für ProgrammiernovizInnen. Darüber hinaus bestätigten die ExpertInnen den hohen Anspruch an Studierende durch metakognitive

Kompetenzen, wie etwa die Abstraktion von Aufgabenstellungen und Handlungsvorschriften, den Transfer von Wissen und Erfahrung auf neue Aufgaben und Programmiersprachen sowie die Entwicklung einer Systematik beim Problemlösen. Die herausfordernde Natur dieser Aufgaben wird durch die Einordnung in die anspruchsvollste kognitive Prozessdimensionen der AKT *Erzeugen* deutlich. Darüber hinaus wird die Rolle des *Metakognitiven Wissens* für die Programmierung sichtbar.

Um die stark abstrahierten Dimensionen und kognitiven Prozesse der AKT insbesondere für die Programmierausbildung zugänglich zu machen, werden in Tabelle 3 die Wissensdimensionen der AKT mitsamt Subtypen neu interpretiert. Darüber hinaus weist Tabelle 4 die Bedeutung der kognitiven Prozessdimensionen nach Anderson und Krathwohl für die grundlegende Programmierausbildung aus, indem Definitionen und Beispiele präsentiert werden. Die beiden Tabellen sind als kontextspezifische Adaption der AKT für die Basisausbildung der Programmierung zu verstehen und können damit als Handreichung zur Klassifikation von Lehr- und Lernzielen in dieser Inhaltsdomäne der Informatik genutzt werden.

Tab. 3: Wissensdimensionen in der Programmierausbildung

| Typen und Subtypen | Definition und Beispiele für die Programmierung |
|---|---|
| **A. Faktenwissen** – Grundlegende Inhalte und Details, die Studierende für die Einarbeitung in ein Thema wissen müssen | |
| A$_A$. Wissen um Terminologie | Fachvokabular (z. B. Sprachelement wie Schlüsselwörter, Escape-Sequenzen, Literale, binäres Zahlensystem, Operatoren, elementare Datentypen, Ausnahmen einer Programmiersprache) |
| A$_B$. Wissen um spezifische Details und Elemente | Fachspezifische Ressourcen und Quellen (z. B. Geschichte der Informatik, des Computers, der Programmiersprachen, Phasen der Softwareentwicklung; Kenntnis zuverlässiger Lehrbücher und Werkzeuge zum Erlernen des Programmierens) |
| **B. Konzeptionelles Wissen** – Übergeordnete Zusammenhänge zw. Elementen als Teile einer übergeordneten Struktur | |
| B$_A$. Wissen um Klassifikationen und Kategorien | Wissen um Programmierparadigmen, Klassen der Zeitkomplexität und algorithmischen Effizienz, Eigenschaften von Algorithmen, Generationen von Programmiersprachen, numerische Datentypen, Chomsky-Hierarchie als Klassifikation für formale Sprachen |
| B$_B$. Wissen um Prinzipien und Verallgemeinerungen | Kenntnisse der Prinzipien von Programmierparadigmen und Programmiersprachen (z. B. Typsysteme, Namensgebung, Funktionen, Parameterübergabe, Speicherverwaltung) |
| B$_C$. Wissen um Theorien, Modelle und Strukturen | Kenntnis der Komplexitätstheorie (Effizienz, Laufzeitkomplexität), Modelle von Computerarchitekturen, Kenntnis von Speichermodellen und Rechnermodellen (z. B. Turingmaschine) |
| **C. Prozedurales Wissen** – Methoden und Regeln für die Anwendung von Algorithmen, Techniken, Strategien | |
| C$_A$. Wissen um fachspezifische Strategien und Algorithmen | Syntax und Semantik von Programmiersprachen, Algorithmen für typische Probleme kennen, Code schreiben, Umwandeln von Dezimalzahlen in Fließkommazahlen (oder anderes Zahlensystem) |
| C$_B$. Wissen um fachspezifische Techniken und Methoden | Konventionen für guten Programmierstil, Methodik zur Analyse von Problemen und Algorithmen, Werkzeuge zur Softwareentwicklung kennen und nutzen, Modellierung und Entwurf von Algorithmen, Algorithmen testen und debuggen |
| C$_C$. Wissen um Kriterien zur Bestimmung der Eignung von Verfahren | Angemessene Verwendung eines Entwurfsmusters oder Algorithmus zur Lösung eines unbekannten Problems (z. B. Auswahl eines problemadäquaten Entwurfsmusters oder Sortierverfahrens) |
| **D. Metakognitives Wissen** – Allgemeines Wissen über Kognition sowie Bewusstsein und Kenntnis der eigenen Kognition | |
| D$_A$. Strategisches Wissen | Wissen um Organisation von Lernprozessen und wann welche Lernstrategien für welchen Zweck sinnvoll eingesetzt werden sollten |
| D$_B$. Wissen über kognitive Aufgaben | Wissen über Bedingungen kognitiver Aufgaben (z. B. nicht-lösbare Probleme, wann zusätzliche Ressourcen zur Problemlösung nötig werden, wann Möglichkeiten zum Transfer von Wissen auf neue Aufgaben bestehen, Bewusstsein über systematische Ansätze zur Problemanalyse und -lösung) |
| D$_C$. Selbsterkenntnis | Reflexion persönlicher Stärken und Schwächen, Bewusstsein über verwendete Strategien, Angemessenheit des Selbstvertrauens und der Selbstwirksamkeit, Bewusstsein über Ziele, Motivation und persönliche Interessen, Übernahme von Verantwortung für Lernprozesse |

Tab. 4: Kognitive Prozessdimensionen in der Programmierausbildung

| Kognitive Prozesse und Alternativen | Definition und Beispiele für die Programmierung |
|---|---|
| **1. Erinnern** – Aufrufen von bekannten Informationen aus dem Langzeitgedächtnis | |
| 1.1 Erkennen<br>Identifizieren | Zugreifen auf Wissen des Langzeitgedächtnisses, das mit vorgelegtem Material konsistent ist (z. B. Daten wichtiger Ereignisse kennen: wann der Computer erfunden wurde und von wem) |
| 1.2 Erinnern<br>Abrufen | Abrufen von relevantem Wissen aus dem Langzeitgedächtnis (z. B. Erinnern an Daten wichtiger Ereignisse in der Geschichte der Informatik) |
| **2. Verstehen** – Ableiten von Bedeutung aus Instruktion und Kommunikation | |
| 2.1 Interpretieren<br>Erklären, Umschreiben, Darstellen, Übersetzen | Wechseln von einer Darstellungsform (numerisch, schriftlich) zu einer anderen (verbal) (z. B. paraphrasieren von Literatur, Dokumentationen, Handbüchern; schriftliche Aufgabenstellung mündlich beschreiben) |
| 2.2 Verdeutlichen<br>Illustrieren, Instanziieren | Aufzeigen eines spezifischen Beispiels bzw. Illustration eines Konzepts oder Prinzips (z. B. Stack als Beispiel eines abstrakten Datentyps erläutern) |
| 2.3 Klassifizieren<br>Kategorisieren, Subsumieren | Zugehörigkeit zu einer Kategorie oder Klasse bestimmen (z. B. zu einer Klasse von Datentypen; Integer als primitiven Datentyp klassifizieren) |
| 2.4 Zusammenfassen<br>Abstrahieren, Generalisieren | Zusammenfassen eines allgemeinen Themengebiets oder Motivs, bzw. eines oder mehrerer wichtiger Punkte (z. B. Kernkonzepte der jeweiligen Programmierparadigmen zusammenfassen) |
| 2.5 Entnehmen<br>Schlussfolgern, Extra-/Interpolieren, Vorhersagen | Aus präsentierten Informationen logische Schlussfolgerungen ziehen (z. B. Zustände von zwei Booleschen Variablen logisch verknüpfen und Wahrheitswert ableiten) |
| 2.6 Vergleichen<br>Gegenüberstellen, Abgleichen, Zuordnen | Erkennen von Übereinstimmungen zwischen zwei Ideen, Objekten, Konzepten (z. B. Gegenüberstellen von Datentypen, Datenstrukturen oder Algorithmen; Von-Neumann und Harvard-Architektur vergleichen) |
| 2.7 Erklären<br>Konstruieren von Modellen | Konstruktion eines Modells zu Ursache/Wirkung eines Systems (z. B. unterschiedlichen Speicherplatzverbrauch und Laufzeit von Iteration vs. Rekursion erklären; Von-Neumann-Architektur erklären) |
| **3. Anwenden** – situative Ausführung, Anwendung oder Implementierung von Prozessen, Vorgängen oder Strategien | |
| 3.1 Ausführen<br>Durchführen | Anwenden eines Verfahrens auf eine bekannte Aufgabe (z. B. Dezimalzahlen in Dualzahlen, Oktalzahlen, Hexadezimalzahlen umkodieren) |
| 3.2 Umsetzen<br>Verwenden | Anwenden eines Verfahrens auf eine unbekannte Aufgabe (z. B. Qualitätskriterien und Programmierkonventionen auf eigenen Quellcode anwenden) |
| **4. Analysieren** – Zerlegung von Material in Teile und logische Erklärung durch Klarlegung derer Beziehung, Bedeutung im Gesamtkontext, Organisation oder Charakterisierung | |
| 4.1 Differenzieren<br>Unterscheiden, Orientieren, Auswählen | Unterscheiden zwischen relevanten und irrelevanten bzw. wichtigen und unwichtigen Teilen eines präsentierten Materials (z. B. eine Problemstellung zerlegen und zur Problemlösung relevante Aspekte auswählen) |
| 4.2 Organisieren<br>Zusammenhänge finden, integrieren, skizzieren, zerlegen, strukturieren | Bestimmen, wie Elemente innerhalb einer Struktur zusammenpassen oder funktionieren (z. B. Bestandteile fremden Codes zerlegen und dadurch dessen Funktionsweise skizzieren). |
| **5. Bewerten** – begründete Auswahl aus mehreren Alternativen, Beurteilung durch Überprüfung von Kriterien und Standards | |
| 5.1 Prüfen<br>Koordinieren, Entdecken, Überwachen, Testen | Erkennen von Inkonsistenzen in Prozessen oder Produkten; Erkennen der Wirksamkeit eines Verfahrens während der Umsetzung (z. B. Testen von Algorithmen und Programmen auf Korrektheit und Eigenschaften, Fehlersuche in Programmen, Verantwortung übernehmen für Lernerfolg) |
| 5.2 Kritisieren<br>Beurteilen | Erkennen von Inkonsistenzen zwischen einem Produkt und extern formulierten Kriterien oder Standards; Erkennen positiver und negativer Eigenschaften eines Produkts; Erkennen der Angemessenheit eines Verfahrens für ein Problem (z. B. Beurteilen, welcher Algorithmus bzw. welches Programm das Angemessenste ist, um ein bestimmtes Problem zu lösen; Komplexität von Algorithmen beurteilen) |
| **6. Erzeugen** – Zusammenstellung von Elementen und Einzelteilen zu einem neuen, funktionierenden Ganzen, (Einzigartigkeit, Originalität) durch Planung, mentale (Re-)Strukturierung, Konzeption, Produktion | |
| 6.1 Generieren<br>Hypothesen bilden | Probleme auf neue Art darstellen und dadurch alternative Hypothesen und Möglichkeiten zur Problemlösung aufstellen; Grenzen von Vorwissen und Theorien werden überschritten (z. B. Modellieren von Problemen; Verschiedene neue Algorithmen zur Lösung eines Problems zusammentragen (Iteration vs. Rekursion); Transfer von Erkenntnissen auf neue Probleme und deren Lösung) |
| 6.2 Planen<br>Designen | Bewusste oder unbewusste Ausarbeitung eines geeigneten Verfahrens zur Erfüllung einer Aufgabe, ggf. Festlegen von Teilzielen und Arbeitsschritten (z. B. Modellieren eines Programms; Algorithmenentwurf; Datenstrukturen entwerfen; Schnittstellen entwerfen) |
| 6.3 Produzieren<br>Konstruieren | Probleme auf Arbeitsplan zur Problemlösung verfolgen und damit ein Produkt erfinden oder entwickeln, das den gestellten Anforderungen gerecht wird (z. B. Programmiersprachliche, lauffähige Lösungen für Probleme schreiben; GUIs programmieren; Systematik zur Problemlösung entwickeln) |

## 5  Diskussion

Die Einordnung der Programmierkompetenzen in die Anderson Krathwohl Taxonomie macht den hohen kognitiven Anspruch an StudienanfängerInnen der Informatik allein im Rahmen der Programmierausbildung deutlich. In kaum einem anderen Studienfach (mit Ausnahme anderer Natur- bzw. Ingenieurswissenschaften) werden von NovizInnen in derartigem Umfang konstruktive Leistungen erwartet. Im Vergleich mit z.Bsp. Sprachwissenschaften fällt auf, dass die dem Programmieren entsprechenden kognitiven Prozesse *Generieren*, *Planen* und *Produzieren* erst beim Verfassen eigener Hausarbeiten, oder gar der Konstruktion einer fiktiven Sprache (z. B. Na'vi, Elbisch, etc.) erforderlich werden. Letztere Aufgabe wird Studierenden durchaus gestellt, jedoch frühestens im Master-Studium. In der Programmierung hingegen stellt das Schreiben eigenen Codes als Methode zur Problemlösung eine konstruktive Leistung dar, die in den ersten Semestern gemeistert werden soll. Paradoxerweise weisen die erhobenen Daten nur in geringem Umfang *Faktenwissen* und *Konzeptionelles Wissen* auf der Ebene des *Erinnerns* und *Verstehens* aus. Gleichzeitig häufen sich Lehr- und Lernziele in den Dimensionen *Erzeugen* und *Prozeduralem Wissen*. Die explizite Ausformulierung niedriger kognitiver Prozesse dieser Wissensdimension wird häufig ausgespart. Wenngleich das Weglassen aufgrund des inkludierenden Charakters der höheren Dimensionen teilweise zu erwarten war, sollten niedrigere kognitive Prozessdimensionen expliziter formuliert werden. Die fehlende Sichtbarmachung von weniger komplexen kognitiven Kompetenzen könnte eine Ursache für zu hohe Erwartungen an ProgrammieranfängerInnen sein, die u.a. von McCracken et al. [Mc01] im Zusammenhang mit den Schwierigkeiten von ProgrammieranfängerInnen identifiziert wurden.

Dem fehlenden Bewusstsein von Lehrenden für die Klassifizierung kognitiver Aufgabenstellungen kann die AKT in der allgemeinen Form nichts entgegen. Durch den hohen Abstraktionsgrad wird zwar eine Gültigkeit für alle Fachgebiete erreicht, doch es können leicht missverständliche Interpretationen erfolgen, wie das angelehnte GI-Modell zeigt. Die erschwerte Zuordnung der einzelnen Dimensionen ist vor allem der semantischen Uneindeutigkeit der verwendeten Schlüsselworte geschuldet. Einige Schlüsselworte, wie etwa „Konzept", können sogar in die Irre führen, da sie eben nicht zwangsläufig auf *Konzeptionelles Wissen* verweisen. Daher ist eine fachspezifische Variation der AKT, insbesondere für die Informatik unabdingbar. Wie das adaptierte Beispiel für die Programmierausbildung in Tabelle 3 und 4 zeigt, kann die AKT in der Grundlagenausbildung der Programmierung durchaus Anwendung finden, um Programmierkompetenzen zu klassifizieren. Die allgemeinen, stark abstrahierten Dimensionen bedürfen lediglich einer fachspezifischen Interpretation für die Informatik um z. Bsp. in der Entwicklung von Curricula,(adaptiven) Assessments und Items genutzt werden zu können.

## Literaturverzeichnis